\documentclass[aps,prfluids,groupedaddress,longbibliography]{revtex4-2}
\usepackage{graphicx}
\usepackage{XiGrpCommon}

\graphicspath{{figs/}}

\usepackage{CJKutf8}

\renewcommand{\RevisedText}[1]{{{#1}}}

\begin{document} % wordcount
% \preprint{APS/123-QED}

\begin{CJK*}{UTF8}{gbsn}
\CJKtilde 
\CJKindent 

\title{Nonasymptotic elastoinertial turbulence for asymptotic drag reduction}
\author{Lu Zhu (朱路)}
\author{Li Xi (奚力)}
\email[corresponding author, Email: ]{xili@mcmaster.ca}
\homepage[]{https://www.xiresearch.org}
\affiliation{Department of Chemical Engineering, McMaster University, Hamilton, Ontario L8S 4L7, Canada}
\date{\today}

\begin{abstract}
Polymer-induced drag reduction is bounded by an asymptotic limit of maximum drag reduction (MDR).
For decades, researchers have presumed that MDR reflects the convergence to an ultimate flow state that is not further changed by polymers.
Our simulation shows that,
as drag reduction converges to its invariant limit, the underlying dynamics continues to evolve with no sign of convergence.
The stage of asymptotic drag reduction is not represented by any single flow state, but encompasses states with varying dynamical patterns, all of which are partially sustained by polymer elasticity.
\end{abstract}


\maketitle
\end{CJK*}

% \begin{document} % wordcount

\section{Introduction}
A small amount of polymer additives can cause substantial drag reduction (DR) in turbulent flows~\citep{Virk_AIChEJ1975,White_Mungal_ARFM2008,Xi_POF2019}.
The effect normally increases with fluid elasticity -- higher polymer concentration or molecular weight.
It is, however, bounded by an asymptotic upper limit -- the maximum DR (MDR), whose mean flow measurements are insensitive to changing polymer solution properties.
\RevisedText{For decades, MDR has remained}
% MDR remains
the most coveted problem in this area.
\RevisedText{A full explanation must consistently address the questions of}
% Both its existence --
what keeps turbulence sustained
\RevisedText{at MDR (the ``existence'' question)}
% high elasticity --
and
% universality --
why \RevisedText{MDR} dynamics displays invariant DR
\RevisedText{(the ``universality'' question).}
% -- have puzzled researchers for decades.

\RevisedText{The earliest attempt was the ``elastic sublayer'' model by \citet{Virk_AIChEJ1975} which considers DR effects to be confined in the turbulent buffer layer.
With increasing DR, the buffer layer enlarges but its growth is eventually bounded by the flow domain size, which causes MDR.
Other theories paint different pictures for polymer-turbulence interactions, but they all resort to the same postulation that MDR represents an ultimate flow state where the increasing length scale affected by polymers reaches an upper limit~\citep{Lumley_ARFM1969,Lumley_JPSMacroRev1973,Sreenivasan_White_JFM2000}.
This limit depends on the flow geometry rather than polymer properties, which explains the universality of MDR. 
However, detailed turbulent dynamics at MDR remained largely missing in all early theories~\citep{Xi_POF2019}.}

% Existing theories for MDR all build on the premise that the underlying turbulent dynamics converges to an ultimate flow state which does not evolve further with increasing polymer elasticity~\citep{Virk_AIChEJ1975,Lumley_JPSMacroRev1973,Xi_POF2019}.
% {\color{blue}Early theory by \citet{Virk_AIChEJ1975} proposed that the effect of polymers is concentrated in buffer layer of the turbulent boundary and in conjecture with the Newtonian plug for sufficiently high $y^+$. MDR is reached when the ``viscoelastic'' sublayer expands to the upper bound of the turbulent boundary layer. This theory, however, was soon proved oversimplified, as the Newtonian plug is preserved only under a low elasticity~\cite{Warholic_Hanratty_EXPFL1999,White_Dubief_POF2012,Xi_Graham_JFM2010,Zhu_Xi_JNNFM2018}.}

\RevisedText{This gap is closed more recently thanks to the advancement in direct numerical simulation (DNS) and flow visualization techniques, from which two}
% The most
promising
% {\color{blue} MDR theory} leads are two
\RevisedText{yet}
diametrical views
\RevisedText{were} proposed in the past decade.
The first considers this ultimate state to be a form of ``hibernating'' turbulence, which is inherently part of Newtonian turbulence but becomes unmasked by polymer elasticity~\citep{Xi_Graham_PRL2010,Xi_Graham_JFM2012,Pereira_Mompean_PRFluids2017}.
% {\color{blue}
% The first considers MDR to be dominated by a form of weak turbulence that is inherently part of Newtonian turbulence but only becomes unmasked by polymer elasticity.
% In this framework, evolution of near-wall turbulent coherent structures is described as intermittent Active-Hibernating-Bursting (AHB) cycles~\citep{Xi_Graham_PRL2010,Xi_Graham_JFM2012,Graham_POF2014,Pereira_Mompean_PRFluids2017,Whalley_Poole_EXPFL2019,Zhu_Xi_JNNFM2019,Xi_POF2019}.
% }
During hibernation, vortices are too weak to adequately stretch polymer chains and generate large polymer stress. Therefore, hibernating turbulence is minimally impacted and cannot be further suppressed by polymers.
Convergence to this Newtonian 
% {\color{blue} hibernating }
flow state perfectly explains the universality of MDR.
However, the scenario has hitherto not been captured in
\RevisedText{DNS}
% direct numerical simulation (DNS)
where the flow \RevisedText{often becomes laminar at high elasticity}~\citep{Xi_POF2019}.
% laminarizes with rising elasticity~\citep{Xi_POF2019}.

%\begin{figure}
%	\centering				
%	% \includegraphics[width=\linewidth, trim=22px 32px 2px 0px, clip]{fig1_2D.png}
%	\includegraphics[width=4in, trim=22px 32px 2px 0px, clip]{fig1_2D.png}
	% \caption{2D EIT solution ($\mathrm{Wi}=100$): (a) time series of TKE contributions; (b) visualization of a representative instant (colors: $\mathrm{tr}(\alpha)/b$; contour lines: $Q=$~\numrange{0.005}{0.02}).}
%	\caption{A typical 2D EIT instant ($\mathrm{Wi}=100$): (colors) $\mathrm{tr}(\alpha)/b$ for polymer extension; (contour lines) $Q=$~\numrange{0.005}{0.02} for vortex strength.}
%	\label{fig:2deit}
%\end{figure}

\RevisedText{The second believes that classical turbulence, which}
% The second believes that the so-called ``elastoinertial'' turbulence (EIT) is the ultimate state~\citep{Samanta_Hof_PNAS2013,Choueiri_Hof_PRL2018,Chandra_Shankar_JFM2018,Chandra_Shankar_JFM2020}.
% Classical turbulence
is fueled by fluid inertia
%, which we call inertia-driven turbulence (IDT),
and suppressed by polymers
\RevisedText{(referred to as inertia-driven turbulence ``IDT''), is replaced by a new type of turbulence in which elasticity drives instability~\citep{Samanta_Hof_PNAS2013,Choueiri_Hof_PRL2018,Chandra_Shankar_JFM2018,Chandra_Shankar_JFM2020}.
This elastoinertial turbulence (EIT) is easily distinguishable from IDT as its vortex structures align in the spanwise (instead of streamwise) direction.}
% , whereas at 
% {\color{blue}
% Classical viscoelastic turbulence is fueled by fluid inertia and suppressed by polymers. This form of turbulence is called inertia-driven turbulence (IDT)~\citep{Xi_POF2019} which represents a broader classes from the weak form of turbulence, such as edge-state, to the intensive Newtonian turbulence. These types of turbulence share many common features such as quasi-streamwise vortices and streaks. Meanwhile, at
% }
% EIT, elasticity becomes a driving force for turbulence~\citep{Dubief_Terrapon_POF2013,Choueiri_Hof_PRL2018}.
% where elasticity drives, rather than suppresses, flow instabilities.
% EIT is a different type of turbulence with
% {\color{blue}
% EIT shows distinct spanwise vortices and tilted polymer sheet structures. % } %(\cref{fig:2deit}). 
Unlike IDT,
\RevisedText{EIT}
% it
exists even in 2D~\citep{Sid_Terrapon_PRFluids2018,Shekar_Graham_PRL2019,Zhu_Xi_JNNFM2020}.
Although 3D IDT structures
\RevisedText{can}
arise intermittently, it is believed that they will vanish at MDR as the flow converges to a pure 2D EIT state~\citep{Sid_Terrapon_PRFluids2018}.
Since EIT is self-sustaining, the existence of MDR is easily addressed. However, its universality remains an ``elephant in the room''.
An instability driven partially by elasticity is expected to vary with polymer properties. Meanwhile, experiments did show invariant DR in an apparent EIT regime~\citep{Choueiri_Hof_PRL2018}.
The elastic nature of EIT and the invariant mean flow of MDR pose a theoretical paradox~\citep{Xi_POF2019}.

\RevisedText{All existing theories are built on the premise that the converged mean flow of MDR implies its converging underlying dynamics.
For half a century, MDR research is synonymous with finding an ultimate flow state that is no longer influenced by polymers -- different theories only differ in what they identify as the ultimate state.
This study will challenge the underlying premise altogether.
We start by showing that a pure 2D form of EIT, as most recently suggested, cannot explain MDR.
The dynamics of MDR must be 3D and, more importantly, it remains nonasymptotic despite the convergence of mean flow.
Interestingly, both the hibernation- and EIT-based theories are parts of a more complex dynamical picture, which is revealed for the first time here.}

% {\color{blue}This study targets to thoroughly understand MDR by revisiting the existing theories through 2D and 3D numerical simulations. }
% We will show that these seemingly contradictory views are both parts of a larger picture, and, quite surprisingly, we find that behind the convergence of friction drag, the underlying dynamics remains non-asymptotic, which challenges the notion of MDR being a converged ultimate state altogether.
% {\color{blue}EIT, in its pure 2D form, does indeed vary with polymer parameters, which has never been examined before. This finding confirms the expected difficulty of an EIT-based theory for the universality of MDR. We do, however, find asymptotic DR in 3D DNS. Quite surprisingly,}
% behind the convergence of friction drag, the underlying dynamics remains non-asymptotic, which challenges the notion of MDR being a converged ultimate state altogether.

% \section{Methodology}
\RevisedText{\section{Formulation}}
We perform DNS for channel flow under fixed pressure drop. The streamwise ($x$-) and spanwise ($z$-) directions are periodic and $y$-direction is bounded by no-slip walls.
The Navier-Stokes equation is coupled with the FENE-P constitutive equation~\citep{Bird_Curtis_1987}
\begin{eqnarray}
	\label{eq:ns:mom}%
		\frac{D \mbf{v}}{D t} =
		- \mbf{\nabla}p + \frac{\beta}{\mathrm{Re}} \nabla^{2}\mbf{v} 
		+ \frac{2\left(1 -\beta\right)}{\mathrm{Re}\mathrm{Wi}}\left(\mbf{\nabla} \cdot
		\mbf{\tau}_p\right),%
\\
	\label{eq:ns:mass}%
		\mbf{\nabla} \cdot \mbf{v} = 0,%
\\
	\label{eq:fenep:conf}
			\frac{\mathrm{Wi}}{2} \left(
			\frac{D\mbf{\alpha}}{D t} 
			 -
			\mbf{\alpha} \cdot \mbf{\nabla v} - \left( \mbf{\alpha} \cdot \mbf{\nabla v}
			\right)^{\mathrm{T}} \right)
			= -\frac{b}{b+5}\mbf{\tau}_p,
\\
	\label{eq:fenep:stress}%
		\mbf{\tau}_p = \frac{b + 5}{b} \left( \frac{\mbf{\alpha}}{1 -
		\frac{\mathrm{tr}(\mbf{\alpha})}{b}} -\left(\frac{b}{b + 2} \right) \mbf{\delta}
		\right),%
\end{eqnarray}
where $\mbf\tau_\text{p}$ and $\mbf\alpha$ are polymer stress and conformation tensors:
polymer extension $\propto\sqrt{\mathrm{tr}(\alpha)}$.
Velocity, length, and time are nondimensionalized by the Newtonian laminar centerline velocity $U_\text{CL}$, half channel height $l$, and $l/U_\text{CL}$, respectively.
Dimensionless parameters include Reynolds number $\mathrm{Re}\equiv\rho U_\text{CL}l/\eta$ ($\rho$ and $\eta$ are fluid density and viscosity),
Weissenberg number $\mathrm{Wi}\equiv 2\lambda U_\text{CL}/l$ ($\lambda$ is polymer relaxation time), viscosity ratio $\beta\equiv\eta_\text{s}/\eta$ ($\eta_\text{s}$ is solvent viscosity), and finite extensibility parameter $b\equiv\max(\mathrm{tr}(\alpha))$.
Quantities in turbulent inner units~\citep{Pope_2000}, denoted by ``+'',
are scaled by the friction velocity $u_\tau\equiv\sqrt{\tau_\text{w}/\rho}$ ($\tau_\text{w}$ is wall shear stress) and viscous length scale $l_\text{v}\equiv\eta/(\rho u_\tau)$.
For instantaneous quantities denoted by ``*'', instantaneous wall shear stress $\tau_\text{w}^*$ is used for inner units~\citep{Xi_Graham_PRL2010}.
The standard domain size is $L_x^+\times L_y^+\times L_z^+=720\times2\mathrm{Re}_\tau\times 230$ ($L_x^+\times L_y^+=720\times2\mathrm{Re}_\tau$ for 2D).
The friction Reynolds number $\mathrm{Re}_\tau\equiv\rho u_\tau l/\eta=84.85$ and $\beta=0.97$ are fixed; $b=5000$ unless otherwise noted.

Spatial discretization uses a hybrid method combining a TVD (total variation diminishing) finite difference scheme~\citep{zhang2015review} for the $\mbf v\cdot\mbf\nabla\mbf\alpha$ term and a pseudo-spectral scheme for the rest\RevisedText{~\citep{Zhu_Xi_JNNFM2020}}.
Artificial diffusion, which may tamper with EIT solutions~\citep{Gupta_Vincenzi_JFM2019,Zhu_Xi_JNNFM2020,Sid_Terrapon_PRFluids2018}, is not used.
Numerical and temporal resolutions are comparable to recent EIT studies~\citep{Samanta_Hof_PNAS2013,Dubief_Terrapon_POF2013,Sid_Terrapon_PRFluids2018,Shekar_Graham_PRL2019}:
$N_x\times N_y\times N_z=256\times 131\times 142$ and $\delta t=\RevisedText{0.004-0.005}$ for 3D and $N_x\times N_y=1280\times 369$ and $\delta t=0.001$ for 2D.
% Numerical details are found in \citep{Zhu_Xi_JNNFM2019,Zhu_PhD2019}.

\RevisedText{
The balance of turbulent kinetic energy (TKE) $\langle k\rangle_\mathcal{V}\equiv(1/2)\langle\mbf v'\cdot\mbf v'\rangle_\mathcal{V}$ is
\begin{gather}
\begin{split}
\frac{\partial\langle k\rangle_\mathcal{V}}{\partial t}
% +\frac{\partial T_y^k}{\partial y}
&= \langle\mathcal{P}^k\rangle_{\mathcal{V}}
	-\langle\epsilon_\text{v}^k\rangle_{\mathcal{V}}
	-\langle\epsilon_\text{p}^k\rangle_{\mathcal{V}}
\\
&= -\frac{1}{2}\int_{-1}^{1}\left\langle v_x'v_y'\right\rangle
		\frac{\partial\left\langle v_x\right\rangle}{\partial y}dy
	-\frac{2\beta}{\mathrm{Re}}\left\langle
		\mbf{\Gamma}':\mbf{\Gamma}'\right\rangle_\mathcal{V}
	-\frac{2(1-\beta)}{\mathrm{ReWi}}\left\langle
		\mbf{\tau}_\text{p}':\mbf{\Gamma}'\right\rangle_\mathcal{V}
\end{split}
\label{eq:tkebal}
\end{gather}
where terms on the right-hand side are the TKE production by inertia, TKE loss by viscous dissipation, and transfer of TKE to elastic energy, respectively~\citep{Zhu_Xi_JNNFM2019,Xi_POF2019}.
$\mbf\Gamma\equiv(1/2)(\mbf\nabla\mbf v+(\mbf\nabla\mbf v)^\mathrm{T})$;
$\langle\cdot\rangle$ and $\langle\cdot\rangle_\mathcal{V}$ denote $xz$- and volume-average, respectively;
``$\prime$'' indicates fluctuating components (e.g., $\mbf v\equiv\langle\mbf v\rangle+\mbf v'$).
At IDT, polymers suppress turbulence for DR and thus $-\langle\epsilon_\text{p}^k\rangle_\mathcal{V}$ is typically negative, while positive $-\langle\epsilon_\text{p}^k\rangle_\mathcal{V}$ indicates net enhancement of turbulence by elasticity.
}

\section{Results}
\begin{figure}
	\centering
	\includegraphics[width=2.3in, trim=0px 0px 0px 0px, clip]{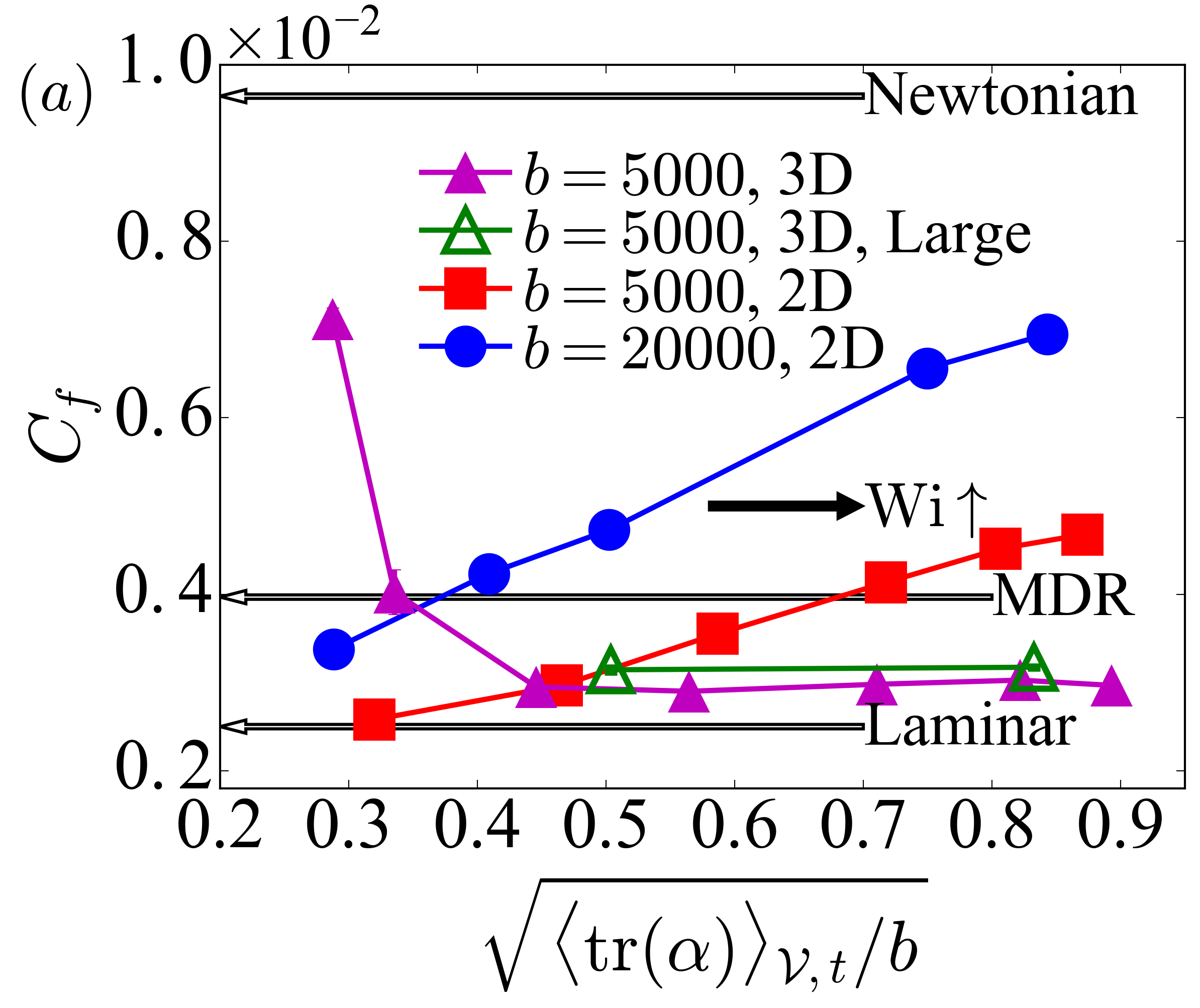}	
	\includegraphics[width=2.5in, trim=0px 0px 0px 0px, clip]{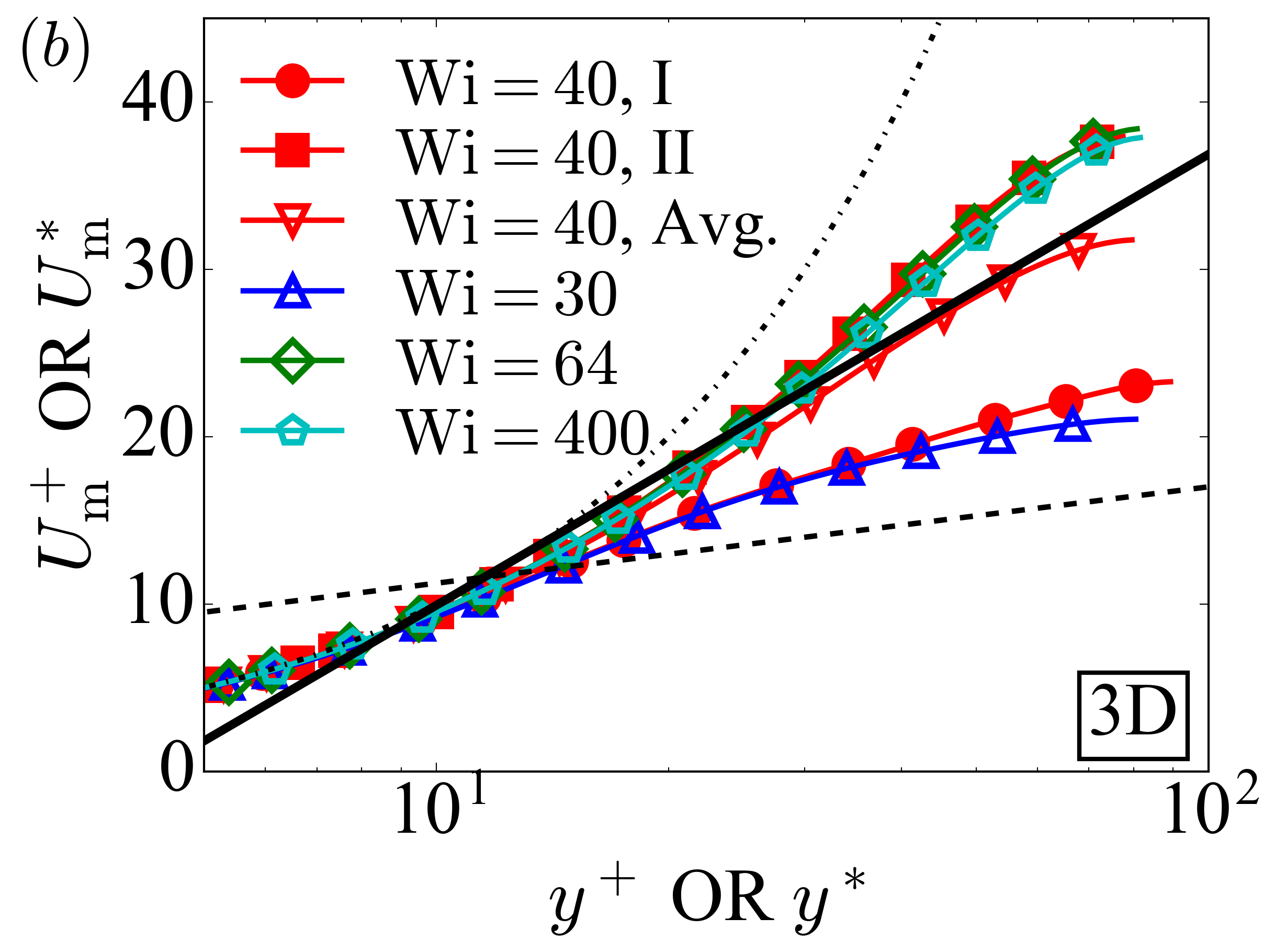}	
	\caption{%
	(a) friction factor (arrows mark Newtonian turbulence, MDR~\citep{Virk_AIChEJ1975}, and laminar values) for 2D EIT and 3D DNS (data provided in SM~\citep{SMref}; the ``Large'' domain is $2L_x^+\times L_y^+\times 2L_z^+$);
	\RevisedText{(b) mean velocity profiles of 3D DNS}
	(with reference lines: (dot-dashed) viscous sublayer $U_\text{m}^+=y^+$; (\RevisedText{dashed}) Newtonian log law $U_\text{m}^+=2.5\ln y^++5.5$~\citep{Kim_Moin_JFM1987}; (\RevisedText{solid}) MDR $U_\text{m}^+=11.7\ln y^+-17.0$~\citep{Virk_AIChEJ1975};
	instants I and II are marked in \cref{fig:trajectory} and \cref{fig:ts}(a)).
	Error bars use block average~\citep{Flyvbjerg_Petersen_JCP1989}.}
	% \caption{\color{blue} { Friction factor (arrows mark Newtonian turbulence, MDR~\citep{Virk_AIChEJ1975}, and laminar values) for 2D EIT and 3D DNS (data provided in SM~\citep{SMref}; the ``Large'' domain is $2L_x^+\times L_y^+\times 2L_z^+$)}}
	% \label{fig:Cf}
	\label{fig:flowstat}
\end{figure}
% \begin{figure*}
% 	\centering
% 	% \includegraphics[width=\linewidth, trim=0px 0px 0px 0px, clip]{fig2_EDTconvergence.png}	
% 	% \includegraphics[width=5in, trim=0px 0px 0px 0px, clip]{fig2_EDTconvergence.png}	
% 	\includegraphics[width=2.5in, trim=0px 0px 0px 0px, clip]{fig2a_mean_ve_SS.png}%
% 	\includegraphics[width=2.5in, trim=0px 0px 0px 0px, clip]{fig2b_mean_ve_wi40.png}	
% 	\caption{{\color{blue} Mean velocity profiles of (a) 2D EIT and (b) 3D DNS, respectively}
% 	(with reference lines: (dot-dashed) viscous sublayer $U_\text{m}^+=y^+$; ({\color{blue} dash}) Newtonian log law $U_\text{m}^+=2.5\ln y^++5.5$~\citep{Kim_Moin_JFM1987}; ({\color{blue} solid}) MDR $U_\text{m}^+=11.7\ln y^+-17.0$~\citep{Virk_AIChEJ1975};
% 	instants I and II are marked in \cref{fig:trajectory} and \cref{fig:ts}(a)).
% 	Error bars use block average~\citep{Flyvbjerg_Petersen_JCP1989}.}
% 	\label{fig:flowstat}
% \end{figure*}
We start with 2D DNS where IDT is precluded and only EIT can exist. 
% From \cref{fig:2deit}, 2D EIT shows distinct thin bands of highly stretched polymers (high $\mathrm{tr}(\alpha)$) \RevisedText{tilted} in an acute angle to the flow direction.
% Vortices are identified by positive $Q$ values \RevisedText{using} the $Q$-criterion~\citep{Hunt_Wray_CTR1988,Zhu_Xi_JFM2019}.
% Spanwise rolls line up near both walls with thinner threads in between.
% This clearly differs from IDT where high $\mathrm{tr}(\alpha)$ regions wrap around mostly streamwise vortices~\citep{Robinson_ARFM1991,Li_Khomami_JNNFM2006,Kim_Adrian_JFM2007,Zhu_Xi_POF2019}.
At $b=5000$, 2D EIT is found for $\mathrm{Wi}$ down to $40$ (\RevisedText{\cref{fig:flowstat}(a)}; also see data in the Supplemental Material (SM)~\citep{SMref}), where polymers are stretched, on average, to $\approx 30\%$ of full extension ($\sqrt{\langle\mathrm{tr}(\alpha)\rangle_{\mathcal{V},t}/b}\approx 0.3$; $\langle\cdot\rangle_{\mathcal{V},t}$ denotes volume and time average).
% and $-\langle\epsilon_\text{p}^k\rangle_{\mathcal{V},t}$. is just above 0.
Its drag, measured by the Fanning friction factor $C_\text{f}\equiv2\tau_\text{w}/(\rho\langle v_x\rangle_{\mathcal{V},t}^2)$, is barely above the laminar level.
Both polymer extension and $C_\text{f}$ increase with $\mathrm{Wi}$.
There is no apparent tendency for $C_\text{f}$ to converge even at the highest $\mathrm{Wi}=800$ where polymers are nearly 90\% stretched.
%Meanwhile, $-\langle\epsilon_\text{p}^k\rangle_\mathcal{V}$ also increases as instability amplifies but seems to plateau as $\langle\mathrm{tr}(\alpha)\rangle_{\mathcal{V},t}\to b$.
%This does not necessarily contradict the increase of $C_\text{f}$, as $C_\text{f}$ depends mostly on velocity fluctuations (Reynolds stress) whereas $-\langle\epsilon_\text{p}^k\rangle_\mathcal{V}$ depends on velocity gradient fluctuations and their correlation with polymer stress fluctuations.
At $b=20000$, the whole $C_\text{f}$ curve is higher, which again rises monotonically with $\mathrm{Wi}$ without convergence.
% The mean velocity profile $U_\text{m}^+(y^+)$ 
% ({\color{blue}\cref{fig:flowstat}(a)})
% declines with both $\mathrm{Wi}$ and ${b}$ to levels much lower than the empirical Virk MDR profile~\citep{Virk_AIChEJ1975}.
Ever growing drag enhancement (DE) with $\mathrm{Wi}$ and $b$ in elastic or elastoinertial instabilities should come as no surprise, which is common in other flow types~\citep{Liu_Khomami_PRL2013}.
It however reveals that, contrary to prior belief, MDR, whose $C_\text{f}$ is invariant with polymer parameters (including $\mathrm{Wi}$ and $b$),
cannot be this 2D EIT state.

By contrast, in 3D flow, DR initially increases ($C_\text{f}$ decreases) with $\mathrm{Wi}$ but later converges.
The last 5 data points in \RevisedText{\cref{fig:flowstat}(a)}, corresponding to $\mathrm{Wi}=64$ to $800$ (see SM~\citep{SMref}), show nearly the same $C_\text{f}$ magnitude.
\RevisedText{Mean velocity profiles (\cref{fig:flowstat}(b))}
% $U_\text{m}^+(y^+)$ profiles
in this range are also inseparable.
% ({\color{blue} \cref{fig:flowstat}(b)}).
The converged DR level slightly exceeds
\RevisedText{the \citet{Virk_AIChEJ1975}}
% Virk
MDR
\RevisedText{asymptote}, which is commonly seen in DNS of comparable regimes~\citep{Dubief_Terrapon_POF2013,Lopez_Hof_JFM2019}. Domain size is one reason.
Two cases ($\mathrm{Wi}=80$ and $400$) tested in a larger domain with doubled $x$- and $z$-dimensions (\RevisedText{\cref{fig:flowstat}(a)}) again have nearly identical $C_\text{f}$. The plateau $C_\text{f}$ level is slightly higher than that of the standard domain%.
\RevisedText{, which is expected given the spatial intermittency and correlation over longer length scales.
We thus expect the mean flow to better match the Virk asymptote in larger domains. Meanwhile,}
% {\color{blue}, most likely owning to the preservation of larger-scale flow structures in large domain simulations which contain intensive turbulent energy.}
% Besides,
the empirical MDR asymptote by \citet{Virk_AIChEJ1975} itself contains uncertainty~\citep{Graham_POF2014,Owolabi_Poole_JFM2017}, which also may not be accurate for lower $\mathrm{Re}$.
% may be attributed to the small simulation domain used.
% Indeed, mean velocity exceeding Virk MDR is commonly seen in 

\begin{figure}
	\centering
	\includegraphics[width=5in, trim=170px 10px 0px 100px, clip]{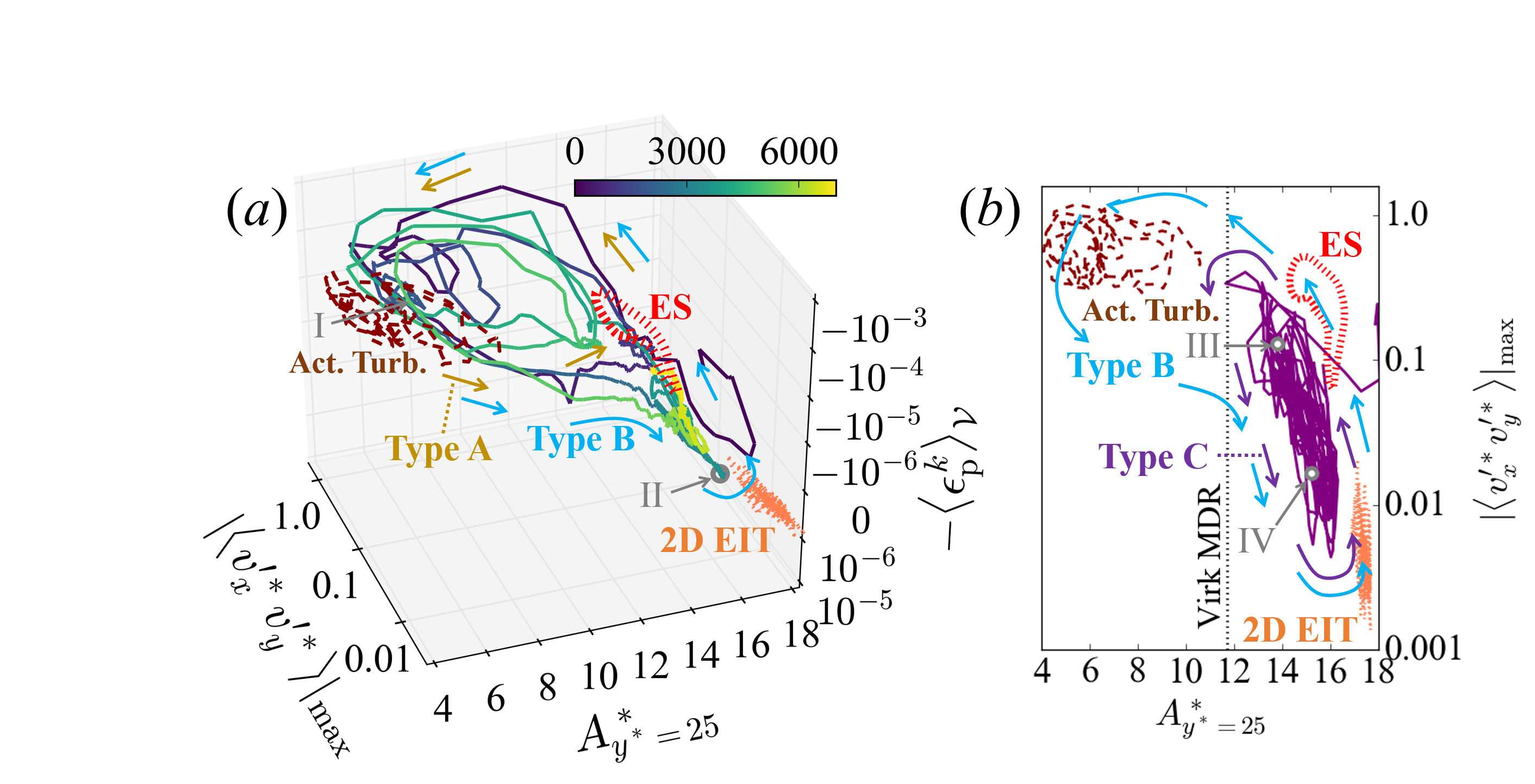}%
	\caption{%
	\RevisedText{Dynamical trajectories of 3D DNS at (a) $\mathrm{Wi}=40$ (solid; color maps to time) and (b) $\mathrm{Wi}=400$ (solid; purple).
	The edge state~\citep{Xi_Bai_PRE2016} and 2D EIT solutions are both at $\mathrm{Wi}=40$; active turbulence is represented by 3D DNS solution at $\mathrm{Wi}=30$.	
	$A^*\equiv y^*(\partial U_\text{m}^*/\partial y^*)$ is the logarithmic slope function and $A^*_{y^*=25}$ measures the local steepness of the $U^*_\text{m}(y^*)$ profile at $y^*=25$~\citep{Xi_POF2019}.
	$|\langle v_x^{\prime *}v_y^{\prime *}\rangle|_\text{max}$ is the maximum instantaneous Reynolds shear stress value (after $xz$-average)~\citep{Xi_Bai_PRE2016,Zhu_Xi_JNNFM2019}.}%
	% A 3D DNS trajectory at $\mathrm{Wi}=40$ triggered by an initial disturbance (solid; color maps to time), in relation to 3D DNS of $\mathrm{Wi}=30$ (dashed), {\color{blue}$\mathrm{Wi}=400$,} the edge state~\citep{Xi_Bai_PRE2016}, and 2D EIT (both $\mathrm{Wi=40}$).
	% $A^*\equiv y^*(\partial U_\text{m}^*/\partial y^*)$ is the local logarithmic slope function~\citep{Xi_POF2019} and
	% $|\langle v_x^{\prime *}v_y^{\prime *}\rangle|_\text{max}$ is the instantaneous Reynolds shear stress profile peak~\citep{Xi_Bai_PRE2016,Zhu_Xi_JNNFM2019}.
	}
	\label{fig:trajectory}
\end{figure}

\RevisedText{The underlying dynamics undergoes transitions between 3 types of cycles.
The $\mathrm{Wi}=40$ trajectory (\cref{fig:trajectory}(a)), which is right before the asymptotic convergence of $C_\text{f}$, shows type~A and type~B cycles.
Both types take the state away from active turbulence -- the dominant form of turbulence of Newtonian flow and lower $\mathrm{Wi}$ characterized by lower mean velocity and higher Reynolds stresses~\citep{Xi_Graham_PRL2010,Xi_Graham_JFM2012,Xi_POF2019}.}
% \Cref{fig:trajectory} shows dynamical trajectories of 3D flow before the asymptotic DR regime.
% The $\mathrm{Wi}=30$ trajectory is mostly confined in the region of active turbulence -- dominant states of Newtonian turbulence with high Reynolds stress and low mean flow~\citep{Xi_Graham_PRL2010,Xi_Graham_JFM2012} ($A^*$ is proportional to mean velocity gradient).gradient
% The $\mathrm{Wi}=40$ trajectory frequently escapes from active turbulence two types of cycles.
Type A cycles visit the vicinity of the edge state (ES) -- an invariant solution that separates laminarizing trajectories from those staying turbulent~\citep{Xi_Graham_PRL2012,Xi_Bai_PRE2016}. These visits are known to be hibernating turbulence~\citep{Xi_Bai_PRE2016,Park_Graham_JFM2015}.
Intermittent active-hibernating cycles occur in IDT even in Newtonian flow and their \RevisedText{statistical weight} increases with $\mathrm{Wi}$~\citep{Xi_Graham_PRL2010,Xi_Graham_JFM2012,Zhu_Xi_JNNFM2019,Park_Graham_PRFluids2018}.
\RevisedText{During type}
% Type
B cycles, which
\RevisedText{are not seen at lower $\mathrm{Wi}$ and reported for the first time here,}
% have not been reported before, emerge at higher $\mathrm{Wi}$:
the flow approaches EIT first and then pivots and returns via the ES region.
\RevisedText{At the high $\mathrm{Wi}$ limit, represented by $\mathrm{Wi}=400$ in \cref{fig:trajectory}(b), the dynamics collapses to a new type C cycle (also never reported before) which appears to bounce back and forth between the vicinities of ES and 2D EIT and strong active turbulence is never triggered.}
% {\color{blue}At $\mathrm{Wi=400}$ (\cref{fig:trajectory}(b)), a new type C cycle occurs which is bounced back and forth between the ES region and the 2D EIT. Unlike the type-B cycles, an IDT is never reached in type-C cycles. }

\begin{figure}
	\centering
	\includegraphics[width=4in, trim=2px 25px 7px 2px, clip]{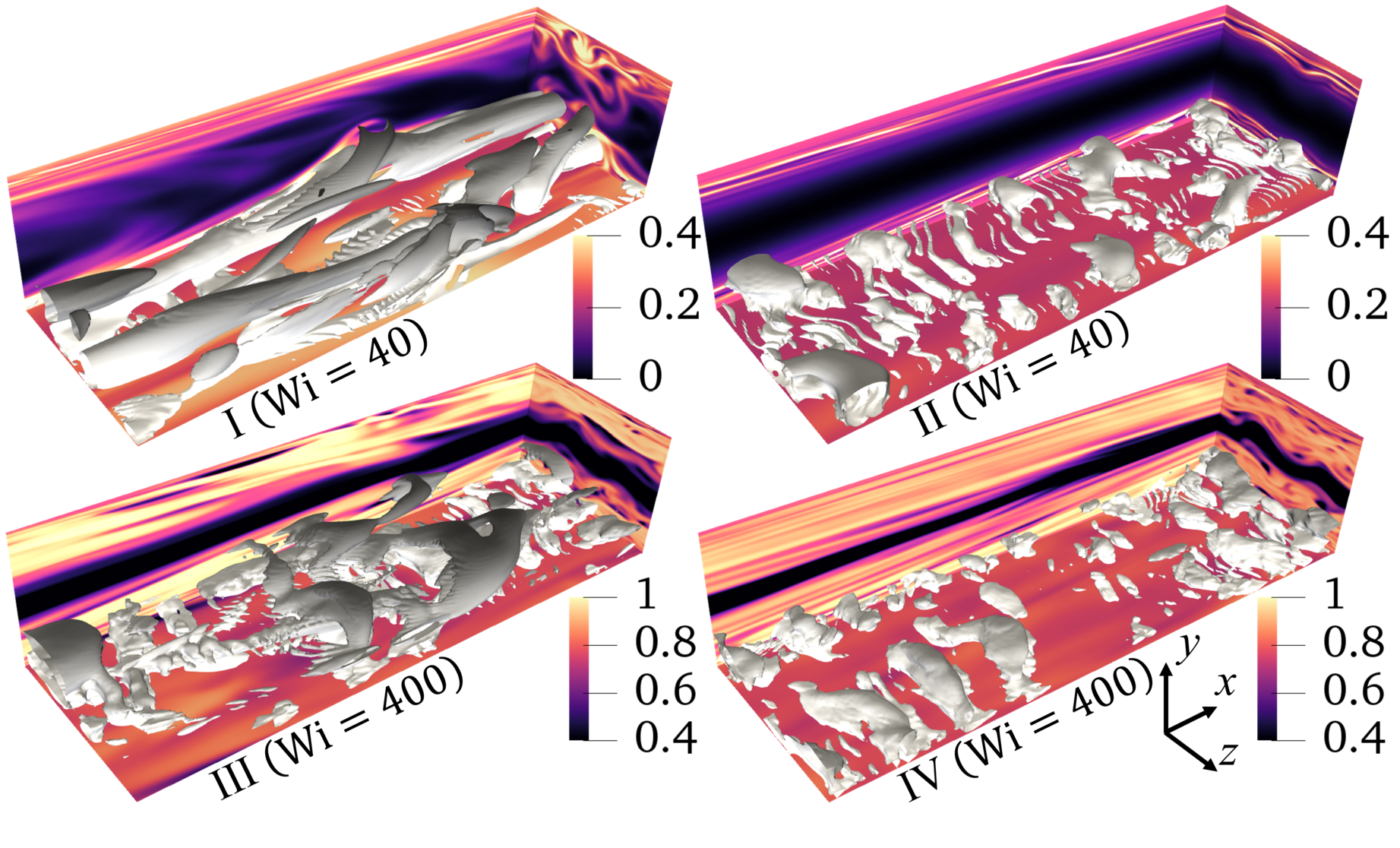}
	\caption{Flow structures of 3D DNS at
	\RevisedText{representative instants as marked in \cref{fig:trajectory,fig:ts,fig:tcorr}.
	Colors map to $\mathrm{tr}(\alpha)/b$. Isosurfaces show vortex configuration using the $Q$-criterion~\citep{Hunt_Wray_CTR1988,Zhu_Xi_JFM2019} ($Q=0.004$; bottom half channel only).}%
	% instants I and II (marked in \cref{fig:trajectory} and \cref{fig:ts}(a)) and III and IV (\cref{fig:ts}(c))
	%(colors: $\mathrm{tr}(\alpha)/b$; isosurfaces: $Q=0.004$ -- bottom half only).
	}
	\label{fig:visual}
\end{figure}

\RevisedText{\Cref{fig:visual} shows flow}
% Flow
structures of different dynamical phases.
% are shown in \cref{fig:visual}.
Active turbulence (instant I) shows classical streamwise vortices
% , although residual EIT structures appear near the wall.
\RevisedText{(some spanwise structures spotted near the wall are likely residues of EIT).
Its instantaneous $U_\text{m}^*(y^*)$ profile (\cref{fig:flowstat}(b)) reflects features of drag-reduced IDT and resembles the time-average profiles of lower $\mathrm{Wi}$.}
Instant II
\RevisedText{represents the EIT region visited during type B cycles and
features trains of spanwise vortex}
rolls spaced by
\RevisedText{thinner}
threads.
\RevisedText{Despite capturing key features of 2D EIT~\citep{Sid_Terrapon_PRFluids2018,Shekar_Graham_PRL2019,Zhu_Xi_JNNFM2020}, it}
% , which, although resembles {\color{blue} \citet{Sid_Terrapon_PRFluids2018}'s 2D EIT},
is not strictly 2D.
Indeed, 3D DNS never fully reaches 2D EIT
\RevisedText{(\cref{fig:trajectory})}, especially at higher $\mathrm{Wi}$ where
\RevisedText{2D EIT}
% the latter
has much higher $C_\text{f}$ (\RevisedText{\cref{fig:flowstat}(a)}).
\RevisedText{Instantaneous velocity of II (\cref{fig:flowstat}(b)) matches the asymptotic DR level of $\mathrm{Wi} =$ \numrange{64}{800} even though at $\mathrm{Wi}=40$ the time average profile is still lower.
Instants III and IV represent two different phases of type C cycles.
Although IV appears similar to II (and thus to 2D EIT), III shows distinct hairpin-like structures in addition to spanwise vortices.
They are probably unrelated with hairpin vortices in Newtonian flow typically observed at higher $\mathrm{Re}$~\citep{Wu_Moin_PoF2009,Shekar_Graham_JFM2018,Zhu_Xi_JFM2019} and possibly result from a fusion between spanwise rolls of EIT and streamwise vortices of IDT (including the ES).}
% The instantaneous $U_\text{m}^*(y^*)$ profile of I ({\color{blue}\cref{fig:flowstat}(b)}) is similar to that of drag-reduced IDT~\citep{Zhu_Xi_JNNFM2018} (compare with the $\mathrm{Wi}=30$ profile in {\color{blue}\cref{fig:flowstat}(b)}), while that of II is closer to 2D EIT ({\color{blue}\cref{fig:flowstat}(a)}).

\begin{figure}
	\centering				
	\includegraphics[width=5in, trim=0px 0px 0px 0px, clip]{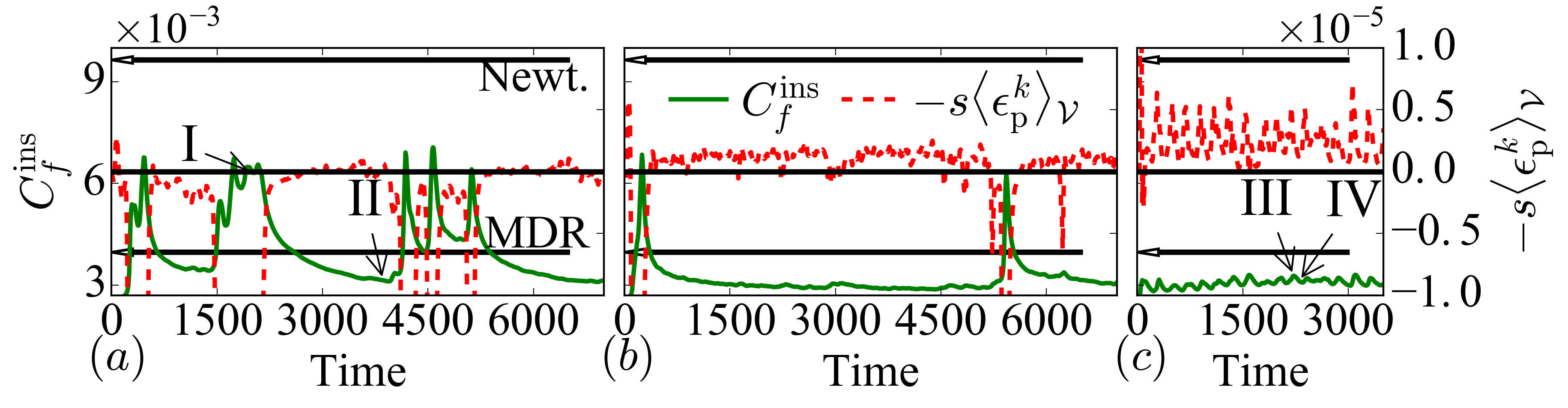}
	%  the "s" for the concersion term are now 1, 3, 5 for Wi=400,80,40
	\caption{Time series of instantaneous friction factor $C_\text{f}^\text{ins}$ and
	\RevisedText{TKE elastic conversion rate $-\langle\epsilon_\text{p}^k\rangle_\mathcal{V}$ (scaled by a factor $s$ for clarity)}
	of 3D DNS: (a) $\mathrm{Wi}=40$ ($s=20$)
	\RevisedText{showing types A and B cycles,}
	(b) $\mathrm{Wi}=64$ ($s=12$)
	\RevisedText{showing type B cycles, and}
	(c) $\mathrm{Wi}=400$ ($s=4$)
	\RevisedText{showing type C cycles.}
	Arrows mark $C_\text{f}$ of Newtonian turbulence and MDR.}
	% \caption{Time series of instantaneous friction factor $C_\text{f}^\text{ins}$,
	% TKE production $\langle\mathcal{P}^k\rangle_\mathcal{V}$ and elastic conversion $-\langle\epsilon_\text{p}^k\rangle_\mathcal{V}$ (scaled by $s$) rates of 3D DNS: (a) $\mathrm{Wi}=40$ ($s=20$), (b) $\mathrm{Wi}=64$ ($s=12$), (c) $\mathrm{Wi}=400$ ($s=4$), and (d) enlargement of (c).
	% Arrows mark $C_\text{f}$ of Newtonian turbulence and MDR.}
	\label{fig:ts}
\end{figure}

% {\color{blue}
% Whether elasticity is a driving or suppressive force on turbulence is determined from the balance of average turbulent kinetic energy (TKE)
% $\langle k\rangle_\mathcal{V}\equiv(1/2)\langle\mbf v'\cdot\mbf v'\rangle_\mathcal{V}$~\citep{Zhu_Xi_JNNFM2019,Xi_POF2019}:
% \begin{equation}
% \frac{\partial\langle k\rangle_\mathcal{V}}{\partial t}
% % +\frac{\partial T_y^k}{\partial y}
% = \langle - v_x'v_y'\frac{\partial v_x}{\partial y}\rangle_\mathcal{V}
% -\langle\frac{2\beta}{\mathrm{Re}}\mbf{\Gamma}':\mbf{\Gamma}' \rangle_\mathcal{V}
% -\langle\frac{2(1-\beta)}{\mathrm{ReWi}}\mbf{\tau}_\text{p}':\mbf{\Gamma}'\rangle_\mathcal{V}
% \label{eq:tkebal}
% \end{equation}
% where the three terms on the right-hand side of \cref{eq:tkebal} are the TKE production by inertia $\langle\mathcal{P}^k\rangle_{\mathcal{V}}$ and TKE losses by viscous dissipation $-\langle\epsilon_\text{v}^k\rangle_{\mathcal{V}}$ and elastic conversion $-\langle\epsilon_\text{p}^k\rangle_{\mathcal{V}}$, respectively.
% $\mbf\Gamma\equiv(1/2)(\mbf\nabla\mbf v+(\mbf\nabla\mbf v)^\mathrm{T})$%
% ; $\langle\cdot\rangle_\mathcal{V}$ denote volume averages, and
% ``$\prime$'' indicates fluctuating components (i.e., $\mbf v\equiv\langle\mbf v\rangle+\mbf v'$; where $\langle\cdot\rangle$ is $xz$-average).
% Positive $-\langle\epsilon_\text{p}^k\rangle_\mathcal{V}$ indicates net enhancement of turbulence by elasticity.
% }

\RevisedText{Instantaneous}
friction factor $C_\text{f}^\text{ins}\equiv2\tau_\text{w}^*(t)/(\rho\langle v_x\rangle_\mathcal{V}^2)$
\RevisedText{at $\mathrm{Wi=40}$} (\cref{fig:ts}(a))
% At $\mathrm{Wi=40}$, instantaneous friction factor $C_\text{f}^\text{ins}\equiv2\tau_\text{w}^*(t)/(\rho\langle v_x\rangle_\mathcal{V}^2)$ (\cref{fig:ts}(a))
shows intermittent dives separated by spikes.
The
\RevisedText{spikes}
reflect returns to active turbulence%
% with strong $\langle\mathcal{P}^k\rangle_\mathcal{V}$ and negative $-\langle\epsilon_\text{p}^k\rangle_\mathcal{V}$
\RevisedText{, which is accompanied by strong}
negative $-\langle\epsilon_\text{p}^k\rangle_\mathcal{V}$
\RevisedText{showing that stretched polymers suppress turbulence.}
Type A cycles display shallow dives (e.g., near $t=4500$) lasting for $O(100)$ time units (TUs). These hibernating intervals still belong to IDT with negative $-\langle\epsilon_\text{p}^k\rangle_\mathcal{V}$ but their $C_\text{f}^\text{ins}$ can approach the MDR level.
Type B cycles show deeper dives lasting for $O(1000)$~TUs with lower $C_\text{f}^\text{ins}$.
They spend most time near EIT where $-\langle\epsilon_\text{p}^k\rangle_\mathcal{V}$ turns positive.

The hibernation-based theory
\RevisedText{conjectures}
that, with increasing $\mathrm{Wi}$, active turbulence diminishes and the flow converges to the ultimate state of hibernation~\citep{Xi_Graham_PRL2010,Xi_Graham_JFM2012}\RevisedText{,}
because the ES would persist as an invariant barrier that blocks
\RevisedText{it}
from laminarization~\citep{Xi_Graham_PRL2012,Xi_Bai_PRE2016}.
% hibernation would persist as the invariant ultimate state while active turbulence diminishes with increasing $\mathrm{Wi}$~\citep{Xi_Graham_PRL2010,Xi_Graham_JFM2012}
Findings here show that
at higher $\mathrm{Wi}$, the flow can move past the ES and EIT steps up as the second line of defense to keep turbulence sustained.
The $\mathrm{Wi}=40$ case, where $C_\text{f}$ has not yet converged, is in the middle of this transition.
At $\mathrm{Wi}=64$, where $C_\text{f}$ starts to plateau, type A cycles disappear (\cref{fig:ts}(b)) and EIT becomes solely responsible for pivoting laminarizing trajectories. The dives are also significantly prolonged.
If $\mathrm{Re}$ were lower, bypass of the ES could occur before EIT emerges. There would then be a window of $\mathrm{Wi}$ where the flow laminarizes,
\RevisedText{as observed in recent experiments}%
%which explains recent experimental observations
~\citep{Choueiri_Hof_PRL2018,Chandra_Shankar_JFM2020}.

\begin{figure}
	\centering
	\includegraphics[width=2.6in, trim=3px 7px 8px 2px, clip]{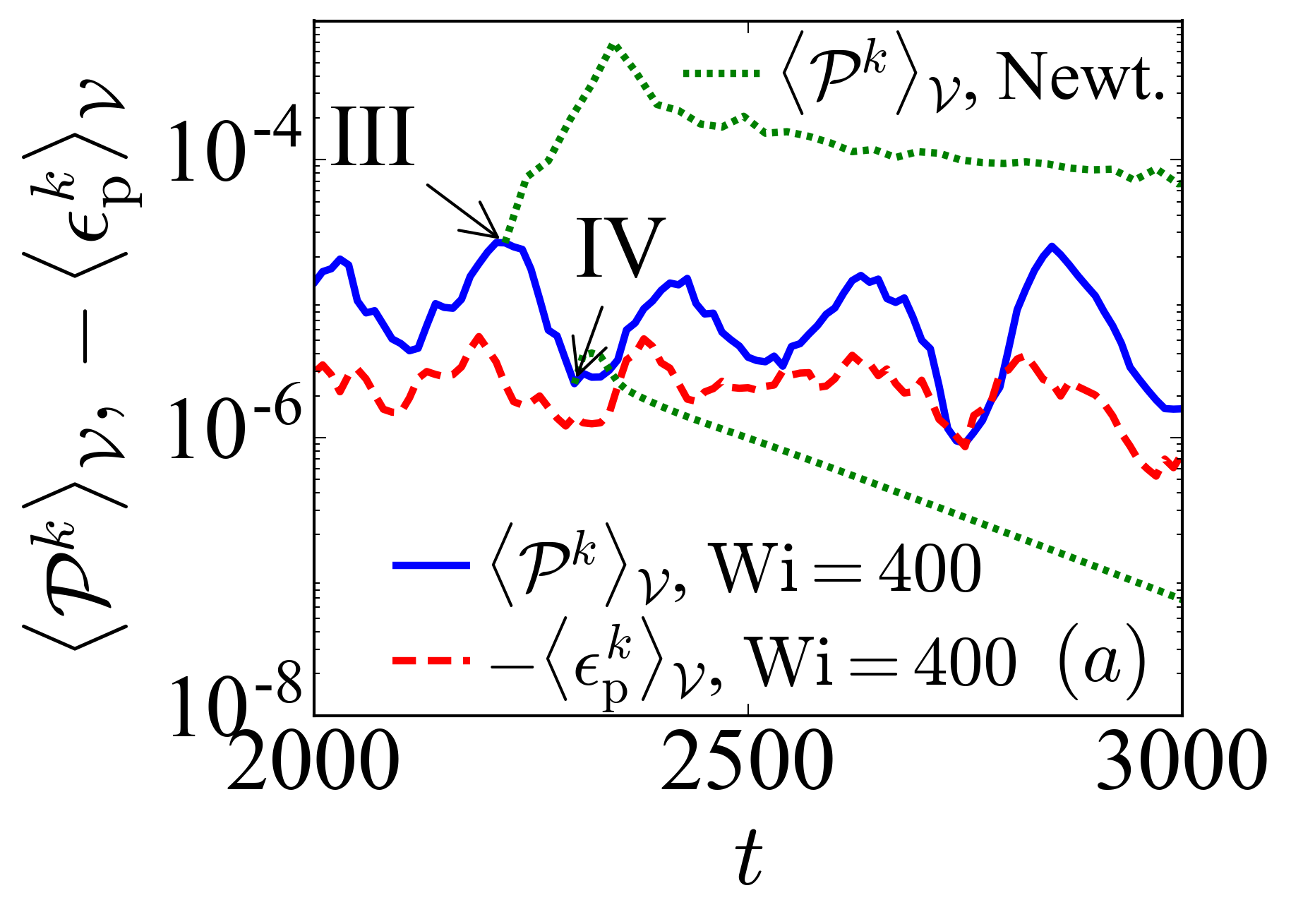}%
	\includegraphics[width=2.5in, trim=3px 7px 8px 2px, clip]{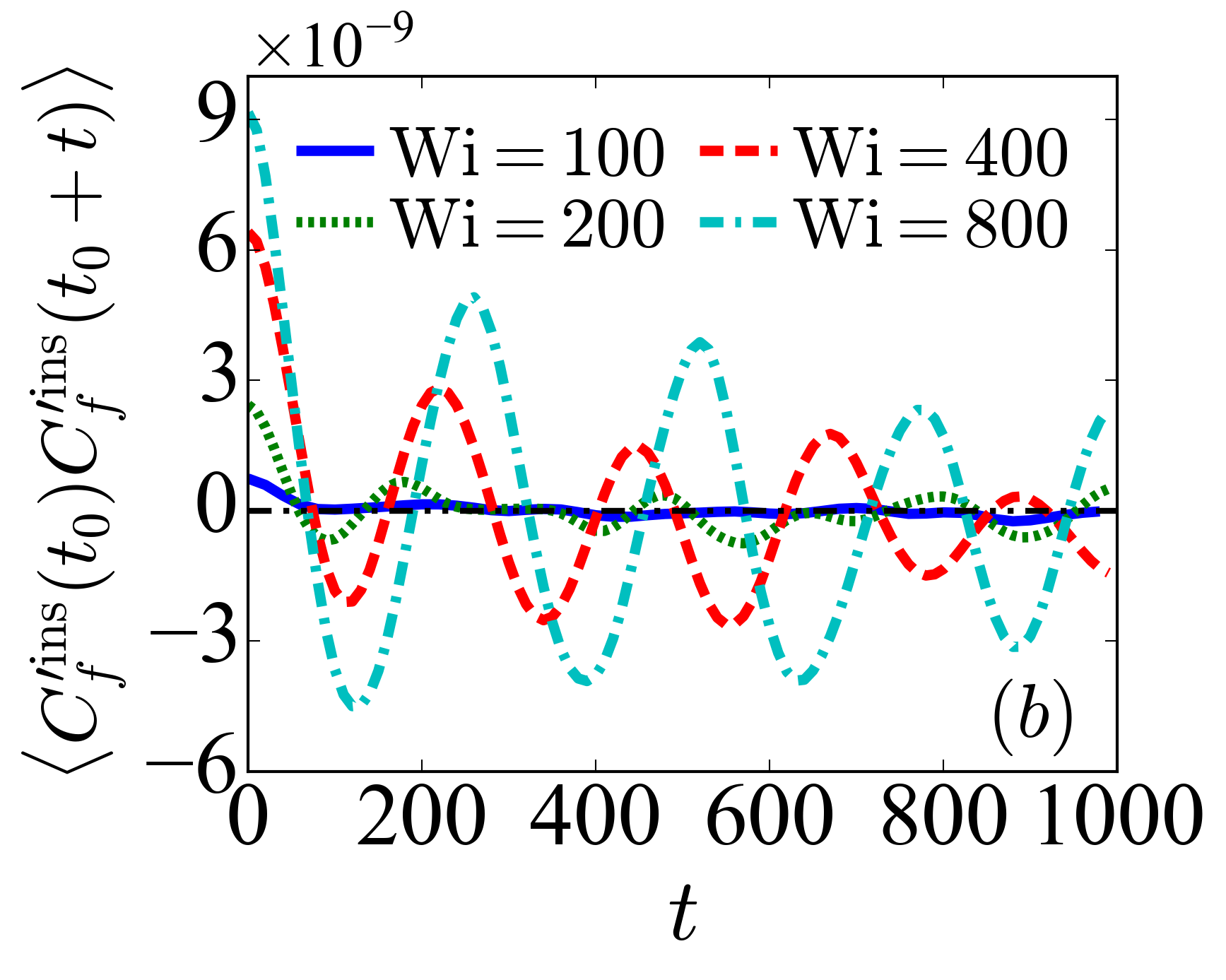}
	\caption{%
	(a):
	\RevisedText{enlargement of type~C cycle (\cref{fig:ts}(c)) time series, with Newtonian shooting trajectories showing the dynamics after removing polymer stress from instants III and IV;}
	% Newtonian shooting trajectories initiated from instants III and IV (marked in \cref{fig:ts}(d));
	% (a): Newtonian shooting trajectories initiated from instants III and IV (marked in \cref{fig:ts}(d));
	(b) time autocorrelation function of $C_\text{f}^{\prime,\text{ins}}$ (3D DNS).}
	\label{fig:tcorr}
\end{figure}

Strikingly, despite $C_\text{f}$ converging from $\mathrm{Wi}=64$ up to $\mathrm{Wi}=800$ tested (\RevisedText{\cref{fig:flowstat}}(a)), dynamical patterns change again for $\mathrm{Wi}\gtrsim 200$.
At $\mathrm{Wi}=400$ shown in \cref{fig:ts}(c), intermittent
$C_\text{f}^\text{ins}$ spikes are replaced by smaller quasi-periodic wiggles.
These 
\RevisedText{type C}
cycles also show much higher frequency than both previous types.
Each cycle (\RevisedText{\cref{fig:tcorr}(a)}) starts with the rise of $-\langle\epsilon_\text{p}^k\rangle_\mathcal{V}$, which quickly sparks a stronger inertial instability, marked by a much higher peak in 
\RevisedText{TKE production}
$\langle\mathcal{P}^k\rangle_\mathcal{V}$\RevisedText{.}
% (in \cref{fig:ts}(c) \& (d), $-\langle\epsilon_\text{p}^k\rangle_\mathcal{V}$ is scaled by $s=4$).
% At highest $\langle\mathcal{P}^k\rangle_\mathcal{V}$ (instant III), distinct hairpin-like structures are observed (\cref{fig:visual}). They seem like a fusion of streamwise vortices of IDT and spanwise rolls of EIT and are likely unrelated with hairpin vortices known in Newtonian turbulence at higher $\mathrm{Re}$~\citep{Wu_Moin_PoF2009,Shekar_Graham_JFM2018,Zhu_Xi_JFM2019}.
% Thinner threads of EIT are also found near the wall.
The net $-\langle\epsilon_\text{p}^k\rangle_\mathcal{V}$ remains non-negative,
\RevisedText{because even at highest $\langle\mathcal{P}^k\rangle_\mathcal{V}$ (instant III) substantial EIT structures persist especially near the wall (\cref{fig:visual}) and $-\langle\epsilon_\text{p}^k\rangle_\mathcal{V}$ is a spatial average of both}
% which is a spatial average of
IDT (negative $-\langle\epsilon_\text{p}^k\rangle_\mathcal{V}$) and EIT (positive $-\langle\epsilon_\text{p}^k\rangle_\mathcal{V}$) structures.
% {\color{blue} In \cref{fig:tcorr}(a), we perform two Newtonian DNS initiated from instants III and IV and show the inertia production  $\langle\mathcal{P}^k\rangle_\mathcal{V}$ as a function of time. Removing polymer stress at moment III will trigger full scale IDT}
Nonetheless, if we remove polymer stress at
\RevisedText{instant III (i.e., Newtonian DNS using its flow field as the initial state),}
% moment III,
full scale IDT will develop (\cref{fig:tcorr}(a)).
% {\color{blue}; the upward shooting trajectory}), indicating that turbulence thereat is fundamentally driven by inertia and elasticity suppresses the further growth of IDT.
\RevisedText{By contrast, during}
% During
the quiescent phase
\RevisedText{(instant IV),}
% (moment IV), flow structures are more typical of EIT (\cref{fig:visual}; compare with moment II) and
removing polymer stress would instead cause laminarization\RevisedText{.}
% (\cref{fig:tcorr}(a)).
% {\color{blue}; the downward decaying trajectory}).
EIT is thus required to trigger the next cycle and sustain turbulence.

\Cref{fig:tcorr}(b) shows the time autocorrelation function of the fluctuating component of $C_\text{f}^\text{ins}$.
At $\mathrm{Wi}=100$, the dynamics remains chaotic and the correlation decays monotonically. Oscillatory patterns appear at $\mathrm{Wi}=200$ as the first sign of quasi-periodicity.
The dynamics continues to evolve afterwards.
% The dynamics does not converge after reaching these type C cycles either.
The amplitude of oscillation increases with $\mathrm{Wi}$ up to the highest level of $800$ tested while its frequency decreases.

\section{Conclusions}
In summary,
we find that for a wide range of $\mathrm{Wi}$ ($64$ to $800$) where DR has reached its asymptotic level, the underlying dynamics never converges. Intermittent dynamics involving bursts of active turbulence persists up to $\mathrm{Wi}\approx 200$, after which quasi-periodic oscillation dominates. The pattern of oscillation continues to evolve with $\mathrm{Wi}$.
As such, the long-standing premise of MDR being a converged ultimate state must be reexamined and continued search of such a single flow state would be misguided.
% continued search for a single ultimate state of MDR would be misguided.

The earlier hibernation-based theory describes the dynamics accurately up to a moderate $\mathrm{Wi}$ level, but direct hibernation is later bypassed, exposing EIT as the second barrier before laminarization.
The EIT-based theory also appears over-simplistic. The flow does not converge to a pure 2D form of EIT. The phenomenological MDR encompasses an ensemble of different self-sustaining turbulent processes which all share the same mean drag.
One may, of course, generalize the concept of EIT to cover all types of turbulence where elasticity is essential for sustained instability, so that all different states within MDR fall under this umbrella. However, that would not erase the nonasymptotic nature of MDR (at least in the low-$\mathrm{Re}$ regime studied), which has never been revealed before.
% . The classical inertia-driven mechanism always plays a part.}

In a larger simulation domain, both intermittent ($\mathrm{Wi}=80$) and quasi-periodic ($\mathrm{Wi}=400$) dynamics are again observed while $C_\text{f}$ stays invariant.
In those solutions, characteristic structures of different dynamical phases, e.g., active turbulence and EIT, are seen to coexist in different regions of the domain, suggesting that temporal intermittency observed in small domains would map to spatial intermittency in experimental flow scales.

% In summary, we conclude that MDR cannot be, as commonly believed, a simple converged EIT state.
% Rather, EIT is non-asymptotic with increasing DE as elasticity increases.
% Asymptotic DR is observed only in 3D
% flow where $C_\text{f}$ is invariant over a wide $\mathrm{Wi}$ range. Its underlying dynamics is, however,
% intrinsically intermittent (i.e., not confined in any single state) and non-asymptotic (i.e., evolving with $\mathrm{Wi}$).
% Such dynamics echoes the earlier AHB framework except that at high $\mathrm{Wi}$, EIT gradually replaces the lamIDT boundary as the main barrier shielding the flow from laminarization.
% Recurrent bursts of IDT are perpetual and also indispensable, which prevents the flow from fully converging to the non-asymptotic EIT.
% At our highest $\mathrm{Wi}$, such bursts are quickly suppressed by polymer stress each time they occur, leading to a dynamical balance between EIT and IDT.
% Finally, the proposed scenario does not contradict the experiments by \citet{Choueiri_Hof_PRL2018}: however, their resurgent turbulence stage, which was believed to be EIT, likely contains recurrent IDT.
% Indeed,
% growing intermittency with polymer concentration is noticeable in experimental flow patterns 
% despite the converging $C_\text{f}$~\citep{Lopez_Hof_JFM2019}.

% \end{document} % wordcount

\begin{acknowledgments}
% \emph{Acknowledgments} --
The authors acknowledge the financial support from the Natural Sciences and Engineering Research Council of Canada (NSERC; No.~RGPIN-2014-04903) and the allocation of computing resources awarded by Compute/Calcul Canada.
The computation is made possible by the facilities of the Shared Hierarchical Academic Research Computing Network (SHARCNET: www.sharcnet.ca).
Our viscoelastic DNS code is based on the Newtonian code \texttt{ChannelFlow 2.0} (https://www.channelflow.ch) developed by John Gibson, Tobias M. Schneider and co-workers.
\end{acknowledgments}

\bibliography{FluidDyn,General,Zhu,sm}

\end{document}